# The Social Abduction of Science

Eamon Duede[a]

James Evans[ab]

[a]University of Chicago

[a]Knowledge Lab

[b]Santa Fe Institute

## Abstract

The logic of abduction involves a collision between deduction and induction, where empirical surprises violate expectations and scientists innovate to resolve them. Here we reformulate abduction as a social process, occurring not only within individual scientists, but often through conversation between those who understand a particular scientific system and its anomalies—its insiders—and those exposed to alien and disruptive patterns, theories, and findings that could resolve them—its outsiders. These extended conversations between scientists, scholars, and disciplines with divergent backgrounds define a social logic of discovery by catalyzing social syllogisms that yield speculative hypotheses with outsized impact across science. We show how this approach theorizes a number of disparate findings from scientometrics and the science and science, tethering them to powerful interpersonal processes that link diversity with discovery. By reframing abductive reasoning as a social process, we attempt to broker a productive new relationship between science studies' focus on scientific practice and the philosophy of science's concern with propositions in scientific discourse, rendering claims-making through abduction a critical scientific practice, central to scientific change, and increasingly available to science studies through small and large-scale data on scientific communication and deliberation.

# Introduction

The Sociology of Scientific Knowledge and Science Studies emerged in the latter part of the 20th century as a reaction to the high-level sociology of scientific institutions pioneered by Robert K. Merton and his students, which largely avoided the study of scientific ideas (Merton, 1942: 116, 1973). Science studies was also inspired by philosophers of science including Kuhn (Kuhn, 1970, 2011), Lakatos (Lakatos, 1980), and Feyerabend (Feyerabend, 1993) who were themselves reacting to the narrow logical empiricist tradition that had focused on the structure of scientific claims (Ayer, 2012; Carnap, 2014; Popper, 2014), but offloaded processes governing their emergence to psychology (Popper, 2005). Philosophers of science from the first half of the 20th Century focused on the logical, propositional substance of scientific discourse. Mid-century sociologists of science focused on the equilibrium of scientific systems. Together, these gave reactive rise to a radical social studies of science that turned toward the unvarnished phenomena of scientific practice.

This paper is, in part, an effort to broker a new and productive relationship between Science Studies, the Sociology of Science, and the Philosophy of Science. We argue that the production of novel hypotheses and claims represents a critical practice of science, central to scientific success and failure, inscribed in the actions of conference mentions, paper citations, and the conferral of accolades and critique. While novel reasoning may involve a psychological process, occurring within scientists' minds and difficult to identify, here we argue that critical aspects of this reasoning involve a social process that can be observed and a logical process that can be analyzed. Peirce defined abductive reasoning as a critical process underlying novel hypothesis generation, which involves the resolution of anomalies and surprises into plausible, empirically verifiable explanations. In what follows, we show that Peirce's logic of abduction does not merely occur within persons, but often between them as a 'social syllogism', manifest in the juxtaposition of distinct experiences between scientists in conversation. We highlight how these experiences, traced through texts and built on different exposures to patterns and persons can be observed and studied through intimate, interpersonal investigation—like the extended conversation between Thomas Hobbes and Robert Boyle at the dawn of the Royal Society in the 17th Century, as recounted by Shapin and Schaffer's *Leviathan and the Air Pump (1985)*. In this way, we argue that science is abducted by the social by showing how social interaction is required for carrying off the logic of abduction.

Specifically, we re-formulate and re-scale abduction to show that it typically occurs through conversation and collaboration between (1) scientists internal to a particular epistemic culture (Cetina, 2009) tuned through education and experience to recognize empirical anomalies as surprising—its insiders; and (2) scientists external to that culture exposed to alien and potentially disruptive patterns, theories, and findings from other areas of science or the world that could resolve insiders' anomalies—its outsiders.

We show that while the scientific experience of surprise is a generative and motivating force for new hypotheses, insiders of a scientific field who are most likely to be surprised by phenomena that violate theory are least likely to be familiar with patterns that could reformulate theory to make those phenomena unsurprising. Our contrast between insiders and outsiders mirrors a frequently made distinction between different kinds of scientists —"hedgehogs and foxes" (Berlin, 1957) or "frogs and birds" (Dyson, 2009), but here we argue that these are *relative* roles and not essential types: an insider in one region of science may be a valuable outsider to another. Moreover, we show how these roles are essential to the functioning of each other for scientific evolution, each conditioned and constrained by personal experience and social interaction, on the one hand, as well as culturally accepted modes of inference and justification on the other. Finally, we articulate how this reasoning conforms with, but also theorizes a number of recent quantitative findings in scientometrics and the science of science. These findings demonstrate how punctuated scientific changes systematically occur through a process of interpersonal abduction where parties to the conversation come from different fields with distinct scientific imaginaries and epistemic commitments (Cetina, 2009).

Central to what follows is our proposal that by acknowledging and examining how scientists engage directly with arguments and claims *in practice*, the social studies of science can extend their understanding of scientific phenomena to the "claimsmanship" of claim conjecturing, stewarding, and defending that is so critical to scientific transformation and scientists' self-understanding. With some mid-century philosophers of science, we argue that scientific claims and arguments are conditioned and constrained by acceptable modes of inference, as well as by prior claims and the arguments that give rise to them (Carnap, 2014; Hempel, 1965; Lakatos, 1980; Popper, 2005; van Orman Quine, 1976). In this way, exposures to and principled engagements with the claims and arguments of a scientist's milieu are as much a part of practice as giving and receiving scientific credit (Latour and Woolgar, 1979; Merton, 1957). This is underscored by how what can be said, accepted, and ratified in scientific discourse must inferentially accord with the machinery of received claims in order to be intelligible, relevant, and justified. Scientists are also conditioned by social and material experiences that impinge upon these claims (Bloor, 1991; Cetina, 2009; Latour, 1987; Martin, 1991). This incorporation of claims within scientific practice provides a new way to evaluate the practical consequences of scientific institutions, by linking how they assemble and channel different kinds of people with distinct backgrounds to how they structure novel scientific claims, making certain hypotheses possible, others inevitable, and still others unimaginable.

Finally, we will explore the relationship between our approach to abduction and the increasing availability of large-scale data from scientific publications and backgrounds. This data can be used to historically evaluate singular and extended conversations in the history of science. At scale, it can also trace scientific exposures to patterns, events and people through prior publication, citation, and training. We show how data of this kind can allow us to identify the social and cultural provenance of claims. Such traces of a claims' pedigree could facilitate the measurement of subjectivities required to further evaluate conjectures like the Foreman thesis (Foreman, 1971; Jammer and Merzbacher, 1967), which contends that patterns in ambient culture can influence the structure and content of scientific claims. Ultimately, this effort could also enable a narrative and calculus of conflicting subjectivities that could enable the characterization and prediction of disagreements, and their likelihood to transform science and scholarly discourse.

## Philosophy of Science and the Social

The philosophy of science has long considered the potential influence of 'social factors'. Early 20th century philosophers of science were concerned with whether there were rational grounds for what they commonly referred to as 'conventionalism' in science. The looming fear in the wake of overthrown Newtonian physics was that adoption of one scientific framework over another was, at bottom, a matter of mere scientific convention and that collective agreement occurred by an accumulation of shared methodology through disciplinary training and exposure (Lakatos, 1980; Poincaré, 1905).

While philosophers of science accepted the presence of social forces at the heart of the scientific enterprise, they fought against them by gerrymandering a conceptual space in which to preserve, certify, and even 'mummify' (Hacking, 1983) objective, rational criteria for the adoption of one scientific framework over another (Ayer, 2012; Carnap, 1932, 2014; Lakatos, 1980; Popper, 2005, 2014; Schlick, 1974). In this way, progress in science was assumed to be both possible and rationally certifiable. The emergence of verifiability in 1881 and falsifiability in 1935 as matters of widespread concern, turned these scientific descriptions into desideratum by placing many aspects of scientific investigation outside the boundaries of science (Gieryn, 1983) (see Figure 1). Even today, Popper's account insinuates itself not only into the philosophy and social studies of science literature, but into domestic science policy, which has admitted falsifiability as a litmus test for support. Current *National Science Foundation* guidelines for fundable dissertation research in sociology state that not only do successful proposals "tend to be theoretically framed and make clear contributions to sociological theory", but "the strongest proposals have a research design that permits falsifiability so that the PI can be wrong as well as right."[1]

By the mid 1950's, the dominant view was that science and scientific rationality were "demarcated" and wholly distinct from other forms of knowing. Rational features of science were said to only manifest in the 'context of justification', the space of reasons in which scientific claims were validated and secured. This was taken to be the province of philosophy. The context of justification was thought to be distinct from that of 'pursuit' or 'discovery', which was too "psychologistic" to be suitable for philosophical engagement and left to psychology, sociology, or history to describe. The purported distinction between the context of justification and pursuit became the target of Kuhn's work, which identified the normal science of extending rather than falsifying claims as the very activity that distinguished scientific practice from other forms of coming-to-know. This placed a central scientific activity beyond Popper's scope of relevance for whom the falsification of hypotheses distinguished the practice of science. In the *Structure of Scientific Revolutions,* Kuhn tamed the notion of convention by illustrating how science relied on the process of upholding and advancing conventional assumptions, which progressed through violation and the reestablishment of new ones by rising generations of young scientists (Kuhn, 1970).

---

[1] The framers of this requirement conveniently did not also read or take seriously Popper's claims about the impossibility of social science (Popper, 2005).

Within philosophy, a counter-current of interest arose in the historical adoption and rejection of epistemic and theoretical commitments. Kuhn, Feyerabend and others pursued inquiries into processes through which acceptance and rejection advanced scientific knowledge. In *Against Method*, Feyerabend argued that science could not progress without methodological radicals, "epistemological anarchists" that generated *ad hoc*, inconsistent hypotheses and explicitly violated so-called 'principles' of falsification and established scientific method (1993). Special people, like Galileo, who were willing to go against both scientific and religious conventions (and, as it turns out, most scientific and mathematical standards of rigour), represent the kind of person who could facilitate punctuated scientific change, not through scrupulous attention to scientific methodology, but through propaganda, *ad hoc* explanation, and forms of epistemological sophistry. In this way, Feyerabend's position against concrete methods was also against definite disciplines, fields, and other social institutions that teach and reinforce them. In the same way that only novel methods or observations could make advances, social radicals were required to champion them.

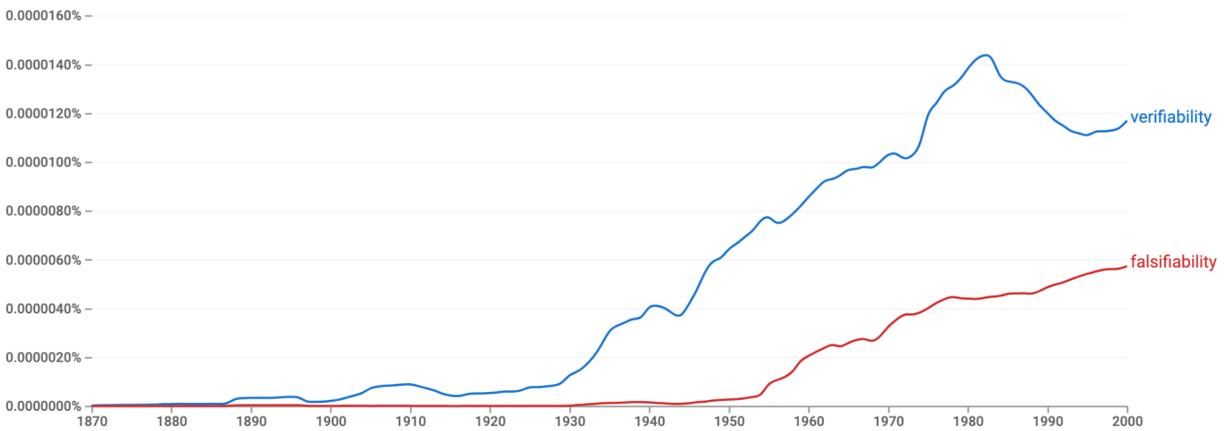

**Figure 1**: Note the emergence of 'falsifiability' in popular discourse in the mid 1950's to counterbalance 'verifiability'. Google n-grams.

Later Hacking, Hull, and a new wave of philosophers began to explore the complex relationship between methods, theoretical insight and the people that held them. Hacking's work on the history of probability suggests that different kinds of statisticians and mathematicians share overlapping views and that paradigm changes may occur less visibly with shifts in a guiding metaphor or interpretation that alters the focus of a field (Hacking et al., 1990; Hacking and Emeritus University Professor Ian Hacking, 2006). Rather than a single totalitarian state of scientific certainty, which could only be overthrown by Kuhn's upstarts[2] or Feyerabend's radicals, Hacking paints a picture of competing scholarly logics that shift as a function of social interaction and intellectual alliance. In *Science as a Process* (2010), Hull goes further to argue that direct competition between schools of thoughts, between the labs that foster them, and selection pressures analogous to mechanisms of natural selection provide a critical motive for scientific publication and advance of one approach over another. Rather than responding to a single, hegemonic epistemic culture (Cetina, 2009), scientists come to be seen as living within some cultures and outside others.

Work of this kind elucidates the homology between intellectual commitments and the social movements that reinforce them. By grappling with social factors at the heart of certification, adoption, and rejection of knowledge claims, philosophical work of this sort came closest to alignment with social and cultural studies of science from the same time period, which tends to focus its attention on scientific practice. Social and cultural studies of science never fully attended to the contributions made to practice by the claims and arguments of scientific discourse, with several, foundational exceptions, which sought to examine both "social" and "technical" aspects (Cetina, 2009; Latour, 1987; Latour and Woolgar, 2013; MacKenzie, 2004). Nevertheless, even these did not focus on issues associated with the interlocking, inferential structure of scientific knowledge and the ways in which this structure interacts directly with other forms of scientific practice.

As a result, neither the philosophy nor social studies of science has taken up the problem of hypothesis generation and discovery as a function of both scientific knowledge and practice, the collision of past justification and present pursuit. As a result, a social studies of science that demonstrates how facts are made still arrives late, having missed the logic at the moment of their genesis, while the philosophy of science continues to regard the issue as philosophically irrelevant, residing beyond the context of justification. Here we argue that a critical examination of abduction, and its typical occurrence as a socially structured conversation can bring these questions together and allow a new productive focus at the intersection of social action and scientific claims.[3]

---

[2] Kuhn acknowledges, endorses and theorizes Max Planck's insight that "a new scientific truth does not triumph by convincing its opponents and making them see the light, but rather because its opponents eventually die, and a new generation grows up that is familiar with it" (Thomas S. Kuhn, 1962, footnote 8).

**Deduction, Induction, Abduction**

19th Century philosopher and scientist Charles Sanders Peirce argued that neither the logics of deduction nor induction alone could characterize the reasoning behind path-breaking new hypotheses in science, but rather their collision through a process he termed abduction. Pierce held that "abduction is the process of forming an explanatory hypothesis"; "studying facts and devising a theory to explain them." For Pierce, this process of "hypothetic inference" forges "all ideas of science" by searching out explanations for surprising new phenomena (Peirce, 1997). In our account, abduction begins as expectations born of theory or tradition become disrupted by unexpected but stable observations or findings. The resulting surprise stimulates scientists to forge new claims that make the surprising unsurprising.

Consider Wilhelm Röntgen's discovery of X-rays at Würzberg in 1895. Since Faraday's production of a glow from voltage in a vacuum, 19th and early 20th century physicists were hungry to understand novel phenomena arising from cathode rays. While implementing a set of well established experimental procedures for producing and studying cathode rays using shielded Lenard and Crookes tubes, but at much higher voltages, Röntgen noticed something unexpected. A barium platinocyanide painted fluorescent screen some feet from his tubes began to glow while the cathode rays were in production (Taylor, 1959). Röntgen wanted to know why the screen glowed under these conditions. Through additional experiments, violations to well-established expectations about cathode rays began to pile up. Röntgen attempted to shield and deflect the cathode rays away from the screen, but they could neither be shielded nor deflected by any of the standard methods—cardboard, aluminum, magnets.

What counted as an anomaly was not how such rays interacted with a barium platinocyanide screen, which even someone unfamiliar with the behavior of cathode rays would have noticed. Rather, Röntgen had strong prior beliefs about how 'cathode rays' interact with magnets and shields, and these beliefs were violated by repeated attempts at empirical interventions. If he was going to persist in attributing the glow to cathode rays, significant restructurings of what was known about them would be required. Qualifications would have to be made about their susceptibility to magnetic fields and shielding under certain conditions, such as when voltages are sufficiently high, or pressure sufficiently low. Seeing what should not occur requires firm expectations about what should. In Röntgen's case, what was not there were cathode rays causing the glow. Lacking other empirical or theoretical patterns on which to draw, Röntgen would simply call them 'X'-Rays, where all that is known about X remains an unknown—a random variable. Here we argue that one way in which anomalous phenomena surprise is by violating prior commitments to principles that determine, in advance, what should have occured; how the world should have been. Röntgen's phenomena conditioned a collision between textbook deduction and induction. The contradiction between what Röntgen knew about cathode rays and what he induced from observations in his lab could have paralyzed him if deduction and induction were more than cartoons of human reasoning in the wild (Hutchins, 1995).

---

[3] More recently, philosophers have taken up questions in the epistemology of discourse with a particular emphasis on testimony. Yet, that literature is concerned to articulate how agents come to acquire (propositional) knowledge by means of conversation. Our concern in this paper is to specify a plausible conversational mechanism for the transmission of claims that can be used to form inferences, regardless of the epistemic status of those claims and resulting inferences.

Consider the deductive and inductive inferences available to Röntgen. Deduction moves from a general to a specific statement such that the logical arrangement of propositions and operators preserves their truth value. In Aristotelian form, the deduction available to Röntgen might go as follows:

1) Cathode rays can be deflected by magnetic fields (general statement)
2) This cathode ray cannot be deflected by magnetic fields (specific observation)
3) Therefore, this cathode ray is not a cathode ray (specific inference)

The major premise (1) states something universal in its truth about cathode rays. The minor premise (2) states something about his specific observation. This deduction might be considered the Khunian deduction which privileges theoretical principles to which Röntgen is committed (1) and concludes (3) with an anomalous, logical contradiction in which a cathode ray is not a cathode ray.[4] To explain this unthinkable outcome, Röntgen might assume something had gone wrong in his experimental setup (2). On the other hand, in the Popperian spirit of refutation, Röntgen could have come to the different explanation that the major premise (1) is false and salvaged the observation that this cathode ray cannot be deflected by magnetic fields because such rays cannot always be deflected by them. Deflection by magnetism simply is not a universal quality of such rays. Yet, Röntgen, himself, insisted that cathode ray deflection had been well established up to that point and that this characteristic of cathode rays will not "be given up except for stringent reasons" (Glasser, 1993).

---

[4] In fact, there are issues with the validity and soundness of this argument. Nevertheless, for illustrative purposes, it should be seen as taking roughly the form of a modus tollens.

Now, suppose Röntgen had no particular commitments and was driven by data. He would not be able to supply the necessary universal statement about cathode rays. Instead, he would rely entirely on inductive inference. An induction moves in the opposite direction, beginning with specific observations and concluding all future observations will follow. Unlike deduction, an induced conclusion is logically uncertain, even if it turns out to be empirically correct. A Baconian representation of the induction available to Röntgen might unfold as follows:

1) In experiment 1, a Crookes tube receives *more* than 5,000 volts and is shielded, but the cathode rays continue to get through (specific observation)
2) In experiment 2, a Crookes tube receives *less* than 5,000 volts and is shielded and the cathode rays do not get through (specific observation)
3) Experiments 1 and 2 are repeated many times with the same outcomes.
4) In experiment 3, a Crookes tube receives *more* than 5,000 volts and a magnet is applied, but the cathode rays continue to persist in straight lines (specific observation)
5) In experiment 4, a Crookes tube receives *less* than 5,000 volts, a magnet is applied and the cathode rays are deflected (specific observation)
6) Experiments 3 and 4 are repeated many times with the same outcomes.
7) Therefore, cathode rays generated at more than 5,000 volts cannot be shielded or magnetically deflected.

This imagined induction is in direct conflict with the deduction above. Based on a deduction that leans heavily on prior knowledge, Röntgen may conclude that cathode rays are not cathode rays; based on induction he may conclude that cathode rays simply behave differently above 5,000 volts. Obviously, Röntgen's thinking in this context was not limited by deduction or induction. These forms of inference are what expose the anomalous phenomenon but are insufficient to resolve it precisely because they operate in the context of justification, not in the context of pursuit. An abduction is a philosophical chimera that like deduction retains the form of a syllogism, like induction is not assured in virtue of logic, but like neither requires familiarity with alien patterns that facilitate an imaginative leap. Consider an abduction that Röntgen might have conceived:

1) The surprising fact that this stable, unshieldable and undeflectable cathode ray is observed.
2) If it were true that 'X' (i.e., unknown) rays existed that were high voltage and could not be shielded or deflected, then having made these observations would be routine.
3) Hence, there is reason to suspect that what I thought were unshieldable and undeflectable cathode rays are actually 'X'-rays.

Premise 2 represents the "abductive leap"; it cannot logically ensure that the conclusion is guaranteed, but there is scant evidence science proceeds through logical assurances. Rather, the first two premises yield the conclusion that what is observed is actually a different type of ray, and the second premise serves as an empirical hypothesis for which additional experimentation and observation are needed to verify or falsify.

The exercise of filling in these hypothetical but plausible templates of reasoning highlights the importance of representations that capture not only the scientific claims scientists make, but also the inferential moves through which they make them. These moves must be perceived as reasonable and probable given both the background theories to which disciplines commit and to the evidence on which these theories are grounded. Moreover, these moves must be plausible not only to the scientist herself, but to the editors, peers, and other audiences who judge them. Inferential moves, and the published scientific claims to which they lead are the public, micro-moves scientists make in their efforts to achieve scientific change. We argue that these claims and moves are critical scientific behaviors and that they should be accounted for and observed, where possible, in combination with other social and cultural data.

Abductive inferences of the form sketched above seek to resolve empirical surprises—collisions between deduced expectations and surprising empirical observations—and, so, lie at the heart of punctuated scientific shifts. Thomas Kuhn calls these cases of discovery-what, where the thing discovered, like oxygen or X-rays, requires novel conceptualization (T. S. Kuhn, 1962). This lies in contrast to cases of discovery-that, which simply involve the observation of the discovered object like an unseen but anticipated element from Mendeleev's periodic table. A true discovery requires discovering both a what and a that. However, with an abduction, the object (a that) is posited before it is understood, and it takes many iterations, people, and backgrounds, often over multiple years, to make sense of the discovery. In this sense, Röntgen's X-rays represent a lame and partial abduction. He was so ensconced in the world of cathode rays that Röntgen couldn't think of anything comparable to these new rays. The new rays were simply unlike cathode rays in very particular ways. He unimaginatively called them 'X-rays', and considered this lack of imagination a personal failing, which, it has been remarked, haunted him for the rest of his life. Nevertheless, 'X-rays' didn't furnish new, generative hypotheses about what properties these new rays might have because generative abductions typically require what we might call a "conversation" between distinct experiences and patterns that rarely reside within the same person.

**Psychic vs. Social Serendipity**

Surprise stimulates scientists to forge new claims that make the surprising unsurprising, and Peirce hinted that the stuff of new hypotheses coalesced by accident within the preconscious mind. He believed that the power of abducing or imagining correct insights about a surprising world arose from "natural evolutionary processes" (Eco and Sebeok, 1983: 17) "developed under the influence of the laws of nature"—"that reason naturally thinks somewhat after nature's pattern" (Peirce, 1929: 269). "Unless man had had some inward light tending to make his guesses…much more often true than they would be by mere chance, the human race would long ago have been extirpated for its utter incapacity in the struggles for existence…" (Muirhead, 1932 Ms. 692). On the basis of Peirce's early work on the psychology of perception at Johns Hopkins University with Joseph Jastrow (1863-1944), he fixed on the degree to which the mind semi-consciously draws on perceptual insight to form hypotheses as insights that come to us "like a flash...putting together what we had never before dreamed of putting together" (Muirhead, 1932, 5.181).

The polymath Peirce experienced more than his fair share of these flashes of subconscious association[5] and intuitive discoveries are well documented in science, such as that of the nineteenth-century German chemist August Kekulé, who claimed to have pictured the ring structure of benzene in 1856, the loop of six carbon atoms structuring organic chemical compounds, after dreaming of a snake eating its own tail (Rocke, 2010). The components of abductive discovery may be planted within the mind by design through education or mentorship. Alternatively, they occur there through serendipity, "discoveries [made] by accidents and sagacity, of things [the observers] were not in quest of" (Merton and Barber, 2004; Walpole, 1754).

---
[5] Peirce describes his own Sherlock Holmes-like sleuthing in his uncanny recovery of a stolen watch and overcoat in a posthumously published essay (Peirce, 1929: 271).

The requirement of serendipity are highlighted in Pasteur's of-quoted maxim "chance favors only the prepared mind", which advises scientists that, in order to make the most of surprising encounters, sufficient preparation is advantageous. But this poses a paradox. The successful scientific mind—the mind prepared to take advantage of a chance encounter—must be simultaneously prepared in two distinct and incompatible ways. On the one hand, that mind must know enough inside the scientific field to be surprised by violations of what is expected. On the other, it must know enough outside the field to imagine one or many reasons why it should not be surprised. This paradox suggests the improbability that abduction to anomaly-resolving hypotheses will occur within a single mind.

Consider abduction as popularized by detectives facing mysterious surprises like Sir Arthur Conan Doyle's Sherlock Holmes. Holmes has twin penchants "for observation and for deduction". He can identify a mystery or anomaly in the micro-data of a crime—a missing shard of glass; a misplaced mound of dirt; the dog that didn't bark (the cathode ray that didn't deflect). Holmes simultaneously fills himself with alien facts and patterns he incorporates into plausible accounts, such that "…when you have eliminated all which is impossible, then whatever remains, however improbable, must be the truth" (Doyle, 1903). It is these alien facts and patterns that, when the dust has settled, remain. But the abductions of Holmes', like Pasteur's lucky scientist, require not one prepared mind, but two. One to understand the situation that gives rise to acknowledgement of and focus on a new mystery or surprise. A second with access to information that would solve or resolve it that lies beyond and is alien to the situation at hand. If this were a single mind, there would be no mystery to solve; no surprise at the scene of the crime. The solution would be assumed. At the limit, this is not a prepared mind, but a Laplacean daemon for whom nothing surprises and everything is predicted.[6]

---

[6] "An intellect [Laplace's daemon] which at a certain moment would know all forces that set nature in motion, and all positions of all items of which nature is composed, if this intellect were also vast enough to submit these data to analysis, it would embrace in a single formula the movements of the greatest bodies of the universe and those of the tiniest atom; for such an intellect nothing would be uncertain and the future just like the past would be present before its eyes." (de Laplace, 1902).

This seeming conflict inspired a logical critique of abduction. If a scientist is familiar with a natural mystery, then their new abductive hypothesis must have already occurred to them, and is therefore neither new nor a discovery at all (Frankfurt, 1958). This conflict disappears if the mind sensitive to conflict between theory and observation and the mind with resources that could resolve that conflict are different minds. If new hypotheses occur through new conversations between people, then abduction as an inference to new explanations ceases to be paradoxical. Despite this critique, in the practice of science, surprises occur and sometimes resolve within the person, as with Röntgen and X-rays or Fleming and penicillin.[7] But the logical critique suggests that abduction will more often occur in conversation across intellects as a social process.

This also follows from the schema that psychologists of science Kevin Dunbar and Fugelsang outline for the process of scientific expansion. They date the beginning of a discovery to the point when a researcher is jarred by the unexpected in observation or experiment, causing self-doubt. This process begins locally but broadens if the researcher and close collaborators conclude the errors are too systematic or persistent to be coincidence. At this point, researchers consider alternative theoretical explanations and may seek the assistance of colleagues whose expertise reaches beyond their own (Dunbar and Fugelsang, 2005). We add to this process that some prominent anomalies, even unresolved, may diffuse and persist in science, passed down across academic generations as problems to solve, greatly increasing the likelihood that someone aware of the problem will converse with someone having the experience or resources available to resolve it. In this way, the significance of many discoveries hinges on surprise in one scientific mind or a community of minds, and a resolving theory or pattern in another brought together through planned or serendipitous conversation.

To illustrate, consider the background knowledge $K_a$ of researcher *a*, which represents all they have read, thought, or imagined. $K_a$ bounds their search for new hypotheses and is some subset of all scientific knowledge $K$—the sum of what is known or imagined by the union of all scientists[8]. When one scientist engages in conversation with another, whether in person, by correspondence, or through reading another's work, the union of their background knowledge necessarily increases what the two can know together[9]. See *Figure 2* for a cartoon of the Petri Dish of conversing scientists. Some scientists share so little knowledge and relevant language that they cannot consummate a conversation (e.g., *a* and *b*). If each scientist possesses a roughly equivalent amount of knowledge, then by maximizing the overlap, they minimize their collective knowledge (e.g., *p* and *q*); by minimizing their overlap, they maximize it (e.g., *x* and *y*).

---

[7] Alexander Fleming followed the surprise of an accidentally infected bacterial sample plate to the cultivation of the penicillin mold to kill bacteria.

[8] $K_a$ not a 'proper subset' of $K$ because individual researchers always hold false beliefs about both the world and what scientists know about the world.

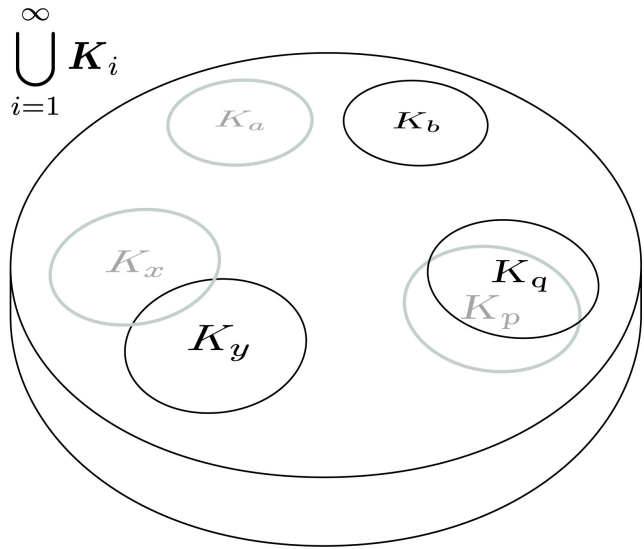

**Figure 2. The Petri Dish of scientific conversations**

Knowledge intersections are not structured randomly, but rather organized into fields and subfields that shape the likelihood that two conversing scientists share stocks of understanding and flows of imagination. Loose versions of this argument shadow calls for increased interdisciplinarity in science and scholarship as the interaction of diverse knowledge bases increases coverage of what is collectively known. Nevertheless, our argument relies not on the existence of interdisciplinary people, but interdisciplinary conversations across people. Moreover, the asymmetry between the two roles involved in abduction—(1) recognizing surprise in anomalies, problems, or conflicts and (2) familiarity with alien patterns that would make the surprising unsurprising—requires a conversation between two types of interlocutors, which stimulate advance at the intersection of surprise and familiarity.

---

[9] We note that what counts as 'knowledge' here is what counts for a given discipline, excluding the possibility that there can be practicing scientists with no knowledge.

## Insiders and Outsiders

Like Pasteur, the astronomer William Herschel imagined a perfect, but impossible scientific reasoner—a scientific Sherlock Holmes—who lived both inside and outside a field: "The perfect observer would have his eyes as it were opened, that they may be struck at once with any occurrence which, according to received theories, ought not to happen; for these are the facts which serve as clews to new discoveries" (Herschel, 1845). The surprise associated with an abduction requires not only "eyes open" to new findings and phenomena, as Herschel suggests, but command of a scientific field and heritage that may only be achievable by insiders. Insiders are scientists raised in the language, lore and research habits of a field. Bourdieu's concept of *habitus* is useful here as it suggests a depth of cultural understanding and practice only acquired through "growing up" within a field. Insiders acquire the *habitus* of their field through education, mentorship, and practice.

Consider the extended apprenticeship of undergraduate, graduate, and post-graduate experiences in a field like physics. This process of induction extends more than a decade and entrains the initiate's expectations to the field's theoretical and technological commitments and associated predictions (Bourdieu, 2004, 2011). This level of understanding is rarely possible without some measure of devotion and discipleship. Kuhn characterized the process of whole-heartedly accepting a field's principles to something like a religious conversion, which involves a gestalt switch to a perspective continuously reinforced by colleagues (Kuhn, 1970). This change in perspective occurs automatically, is not necessarily conditioned by reason, and may be very difficult to reverse because of its implications for identity and membership in the community. Often insiders defend the faith as priests and priestesses, scholastics devoted to the canon of scientific scripture with which they were anointed as scientists. As a consequence, they may be more sensitive to anomalies that violate not only intellectually, but morally, the sacred axioms and treasured understandings of their disciplines. They may come to *feel* that a new finding is inconsistent with their scientific worldview (Keller, 1984).

Herschel's fantasy of the scientist who simultaneously resides inside and outside deviates from reality as, for any limited being, the depth of acquired scientific *habitus* required to sense and identify an anomaly (Bourdieu, 2004) is inversely proportional to the breadth required to have ready access to patterns, methods, and hypotheses that could make that surprise unsurprising. It is the very disciplinary depth, expertise, and *habitus* of a given scientist that may make her least aware of the relevant principles necessary for resolving empirical violations made visible by her expertise. This is because, as Bourdieu might put it, theories are "structured structures predisposed to function as structuring structures"—structuring the very phenomena they describe.

Herschel's ideal scientist, who leverages observed anomalies as clues for the conjecture of new and better theories, anticipates Popper's scientific falsificationist programme. Nevertheless, anomalous observations may rarely be cognitively registered as such. Psychologist Kevin Dunbar has shown that if the goals of science involve confirming and extending existing understanding, it is rare that they come to meaningful explanations of anomalous results (Dunbar, 1993). Scientific minds are often "prepared" in ways that directly obstruct discovery. Consider psychiatrist Robert Heath who failed to recognize evidence of pleasure centers while stimulating the brains of his schizophrenic patients, despite their reports of pleasurable sensation, precisely because he expected them to feel alertness. When some acknowledged alertness, it became the anticipated focus of his study (Baumeister, 2006). Even if pleasure had been acknowledged and investigated, as it was later by Olds and Milner in 1953, Heath did not have the scientific resources to theorize it. In short, anomalies are often set aside, attributed to experimental error or forcibly coerced into agreement with widely accepted tenets of the field.

In some fields commitment to established theory is so strong that empirically plausible alternatives remain not only unarticulated, but largely unimagined. Consider postulations of dark matter and dark energy in astronomy and cosmology (de Swart et al., 2017). Dark matter is currently assumed to make up ~85% of the universe's mass and approximately a quarter of its energy density. Its presence is implied by accepted theories—observed gravitational effects are anomalous unless much more matter is present than can be detected with telescopes or other astronomical instruments. Moreover, in the widely accepted Lambda-CDM model of cosmology, dark energy, also never detected empirically, accounts for 68% of the universe's energy. Yet, in the wake of massive continuing efforts to measure them, there remains a complete lack of empirical evidence for dark matter or energy. Nevertheless, theoretical commitments inspire mainstream physicists imagine not just one form of dark matter, but an entire dark sector of the universe involving diverse forms (van de Bruck and Morrice, 2015; Zumalacarregui et al., 2010), even including "dark life" (Randall, 2017). Commitment to the first principles of relativistic physics school the eye to see what literally cannot be seen. The few who have proposed alternative theories like MOND (Modified Newtonian Dynamics) that do not assume the existence of dark matter or energy, violate hard-core theoretical tenets (Lakatos, 1980) and have not penetrated beyond the extreme periphery (Smolin, 2007). Without outsiders possessing an established body of facts, theories, models and other resources available from another mature science, it is difficult to imagine how a heterodox physicist could acquire the institutional support to seriously explore alternatives.

Scientific disciplines maintain a set of constitutive scientific claims and established methods that limit their ability to sustain the diversity required to solve novel and unanticipated anomalies. Differences between fields limit communication and migration, and so preserve the independence of focus and interest that make abductive discovery between scholars from different fields more likely than those within. These forces include domain-specific jargon (Vilhena et al., 2014), or logical inconsistency and incommensurability between the concepts or forces distinct fields maintain, such as methodological individualism in economics versus the methodological holism in anthropology and parts of sociology. Diverging standards of evaluation may also increase the latent value that members of one field can offer another in resolving an anomaly. Theories can be evaluated based on their accuracy of prediction, internal consistency, breadth in explaining disparate findings[10], parsimony[11], or fruitfulness for generating new questions (Kuhn, 1977; Tria et al., 2014). But different fields place different weights on the criteria that constitute an acceptable explanation. A field may exclude or attack explanations that have been nurtured and thrive in neighboring fields with disjoint epistemic cultures (Cetina, 2009). As a result, conversations between scientists in different fields allow outsiders to propose explanations unconsidered by those inside.

Our focus here has been on how groups of scientists differ, without consideration of individual differences. We do this because access to intellectual resources, available for explanation of a scientific anomaly will more likely differ based on the acquired *habitus* of a scientific field—from the empirical exposures and epistemic cultures within that field—rather than from the idiosyncratic genius, personality, or style of individuals within that field. It is perhaps because Kuhn places the burden of differences in epistemic style on individuals rather than fields (Kuhn, 1977), that his logic of progress misses the social structures that limit access to patterns, models, methods, and epistemic approach, which make conversation between them so fruitful.

---

[10] This is the idea discussed in William Whewell's notion of concillience that "The Consilience of Inductions takes place when an Induction, obtained from one class of facts, coincides with an Induction obtained from another different class. Thus Consilience is a test of the truth of the Theory in which it occurs"(Whewell, 1840).

[11] This has been articulated by William of Ockham as *novacula Occami* (Occam's razor) or *lex parsimoniae* (the law of parsimony) as the problem-solving principle stating states that "Entities should not be multiplied without necessity" (Schaffer, 2015).

**Symmetry of Insiders and Outsiders.**

Legitimate researchers are insiders for *some* scientific domain—the field from which they hold advanced degree(s), the department in which they are employed and teach introductory classes, the journals in which they regularly publish—with mastery that may not carry over to others. Physicist Richard Feynman offers an understated account of his 1959-1960 sabbatical in Max Delbrück's biology lab at Caltech in his memoir *Surely You're Joking, Mr. Feynman* (Feynman, 1985). After developing delicate lab coordination techniques and a rich biological vocabulary, Feynman observed that his "problem-solving algorithm" for physics didn't help much in the Delbrück lab. He made observations others had not yet considered—why do chloroplasts circulate within the cell?—for which there was no ready explanation. With Matt Meselson he focused on ribosomes and with Robert S. Edgar, bacteriophage mutants. After a failed experiment with Hildegarde Lamfrom to identify whether ribosomes from bacteria work in plants or any other living thing, Feynman reflected that "we would have been the first to demonstrate the uniformity of life"...."it would have been a fantastic and vital discovery if I had been a good biologist. But I wasn't a good biologist." He used ribosomes from a prior experiment left so long in the refrigerator that they had become contaminated with other samples and so the experiment did not work (Feynman, 1985). Nevertheless, his work with Edgar was picked up as supporting evidence in a major 1961 *Nature* paper by Crick, Watson and colleagues on the generality of the genetic code for proteins (Crick et al., 1961).

More importantly, spending a year immersed and manipulating molecular "machinery", however amateurishly, with the guiding (if philosophically dubious) view that "there is nothing that living things do that cannot be understood from the point of view that they are made of atoms according to the laws of physics" (Feynman et al., 2005), led to a novel insight. In a 1960 talk he delivered at the annual American Physical Society "There's plenty of room at the bottom" (Feynman, 1960), Feynman introduced the idea of nanotechnology by posing the problem of manipulating and controlling things on very small scales (Mauro et al., 2018). His outsider status was the thing that allowed him to recognize the components of molecular biology as building blocks that could be used in arbitrary physical construction.

Every insider from one field is an outsider to others. And every scientific agent who appears an outsider, sharing neither language nor norms of a field is either a baby or a barbarian—a newcomer to science or an insider from some distant, alien field. And this outsider perspective, if sustained, as Feynman's was through a year-long sabbatical, can cast fresh light on old patterns and problems, or cast a long shadow on old commitments and expectations. A consequence of this phenomenon has been investigated at scale by Azoulay, Fons-Rosen and Zivin in a 2019 *American Economic Review* article titled "Does Science Advance One Funeral at a Time?" (Zivin et al., 2019). Here, they built a dataset of 452 eminent life scientists who died prematurely (e.g., car crash, cancer) to examine whether their deaths conditioned growth, stasis or decline in their subfields. The death of star scientists lowered the barriers that held back relevant ideas from outsiders, whose new articles in the subfield brought new, highly cited perspectives on established problems. This outsider influx did not occur, however, when the star's former collaborators and protegés—priests of the established order—continued to hold editorial influence or the reins of relevant biomedical funding. This pattern suggests that conservative forces resist abductions which might have been made by outsider scientists in conversation with a field's established problems. Nevertheless, social resistance eventually gives way, if sometimes only through death.

In works of sociological self-help that depict how social researchers might enhance their abductive imagination, authors repeatedly argue for the criticality of cultivating diverse theoretical imagination. Stefan Timmermans and Iddo Tavory describe how abduction "depends on the researcher's cultivated position. The disposition to perceive the world and its surprises…is predicated on the researcher's biography as well as on an affinity and familiarity with broader theoretical fields…on the scope and sophistication of the theoretical background a researcher brings to research" (Stefan Timmermans and Iddo Tavory, 2012). Randall Collins suggests that if "thinking is a conversation with imaginary audiences" then "high degrees of intellectual creativity comes from realistically imagining existing or prospective intellectual audiences" (Collins, 1987). Boltanski and Thevenot claim that "innovation tends to occur where more than one standard is at play—more than one order of worth" (2006). But in contrast with the intention of these arguments, the most natural method to cultivate a distinctive theoretical biography, imagine different intellectual audiences or maintain alternative standards is to be a different person, an outsider.

The persistent difference between field insiders and outsiders lies in the disunity of science (Gieryn, 1999; Stump and Galison, 1996). Scientific fields represent a patchwork of partially overlapping epistemic cultures (Cetina, 2009), each holding their own doctrine of concepts, assumptions, and facts, linked by epistemic principles forged by their Nobel Prize winners, Fields Medal winners or otherwise beatified saints. Ordained priests—presidents of professional associations, department chairs, editors of journals—manage cultural boundaries by discerning orthodoxy and declaring heresy. Terminology that facilitates communication within a distinct scientific community requires translation for outsiders (Vilhena et al., 2014). If differences lie deeper in incommensurable concepts, problems, and methods, parsimonious translations may be impossible. Despite purported "physics envy" among economists (Lo and Mueller, 2010; Mirowski, 1999) who sometimes match or exceed the mathematical formalism routine in the physical sciences, their commitments to methodological individualism and the iid assumption (e.g., that agents are identically and independently distributed), make much physical formalism, as that relating to fields and particulate interactions irrelevant (Li et al., 2020). As a result, expertise from one domain unevenly carries over to others, making it unlikely to canonize in multiple fields. None but Marie Curie has won a Nobel Prize in two natural sciences.

Overlaps between domains, points of contact and objects of shared reference make communication between some groups much more likely than others. Experimental and theoretical physicists engage with one another through phenomenologists who sit between them, translating theory into hypotheses and findings into theory. Even when communication is difficult, these fields are sufficiently entrained to the same phenomena that they will exchange machinery, protocols, and findings in scientific trading zones (Galison, 1997) even if their interpretation of these boundary objects is in conflict (Star and Griesemer, 1989). This suggests that even if scientists reside so far apart from one another in cultural space that fluid communication is impossible, pidgin languages constructed for the purpose of trade and the circulation of scientific objects, including methods, protocols and publications facilitate connection between insiders and near outsiders. More improbable communications between certain fields facilitates the diversity and separation that, in turn, enabled differentiation and quasi-independent growth of distinct justifications, techniques and heuristics for research, as also attention to distinct patterns and supposed problems in nature and data. And it is these swirling pools of difference that serve as reservoirs of inspiration to be drawn upon by outsiders to resolve the most vexing, theory-resistant puzzles of insiders.

**Progressive Dilemma and its Social Resolution**

Within a discipline, the first attempted explanation of anomalous phenomena will often be to explain it away, casting doubt on the experimenter or experimental procedure and to resist the finding or idea (Barber, 1961). The first serious engagement will typically draw upon first principles from that discipline, as in cases of Röntgen, physicists of dark matter, and theorists of dark life. Otherwise, their accounts of those phenomena would themselves be instances of disciplinary violation and defection. As such, it is unlikely that a single scientist, brought up in the *habitus* of a single field, will be sensitive enough to her field's principles and commitments to detect anomalies while simultaneously prepared with the heterogeneous intellectual resources to resolve them with theories, patterns, or logic alien to it. In order for the science of a discipline to evolve, expand, and change in its generality, challenges to that discipline's principles must arise and these are likely sourced from encounters with anomalous phenomena perceived as violations of what is expected, such that what is expected becomes revised to accommodate the phenomena. This revision appears particularly unlikely to issue from a disciplinary insider precisely because it is the insider's disciplinary commitment that made the phenomena visibly anomalous in the first place. This creates a *Progressive Dilemma* in which progress to resolve anomalies is not available inside the field, requiring engagement with outsiders whose knowledge and experience are themselves anomalous. We argue that this philosophical puzzle cannot be solved within one philosophizing mind, but requires sociality and communication between diverse minds.

If an anomalous phenomena cannot be explained with resources within the field, scientists must either ignore it or acknowledge their cluelessness. Either the discipline's principles are lacking, or there is something about those principles the scientist is lacking. If the latter, then collaborations with other insiders may be sufficient to show how the 'anomalous' phenomena was to be expected, representing a consolidation of disciplinary principles. In cases where the principles, themselves, are insufficient, progress will require novel principles which may conflict with established thought.

The common mechanism we propose in this paper is the *social syllogism*, an abduction carried out across individuals. An insider's set of disciplinary principles makes observed phenomena inconsistent with what is deducible from theory or extensible from accepted observations. In order to resolve the inconsistency, new claims must be added to the principles of the discipline which are, themselves, in conflict or inconsistent with established principles. The *progressive dilemma* makes it unlikely (though not impossible) that these new claims will issue from within the discipline. But this fictive dilemma is routinely resolved. Science as a whole lacks ontological, theoretical, and epistemological unity, so new concepts, methods or phenomena that could resolve one field's dilemma may be readily available from and supplied by others. Conversation with outsiders enables the forging of experimental claims positioned to transform the anomalous into the obvious.

These insights are anticipated in the sense-making approach to organizational cognition where individuals attempt to make sense of the game they are playing from situated contexts (Weick, 1995). This perspective gave rise to the provocative, "irrational" garbage-can model of decision making in which problems, solutions and social agents are only loosely coupled organizational streams, which only randomly and unexpectedly combine to facilitate problem solution (Cohen et al., 1972). Here we argue for an anti-correlation of problems and solutions in science, where the scientific agents that can identify anomalies and generate problems perceived as interesting to like-minded colleagues are systematically the least likely to hold the alien solutions required to solve them. Unlike in business, where much of the garbage is kept in the can; in science, the garbage bag has exploded and epistemic repulsion limits the potential for the most productive collisions.

Our theory of collective abduction also theorizes a number of recent quantitative findings in scientometrics and the science of science (Fortunato et al., 2018). Recent work on the transmission of knowledge and influence emphasizes the importance of place for catalyzing unexpected collaborations that foment scientific change. One study (Wuestman et al., 2019) observed that scientists co-located within an institution are more likely to cite one another's work, but also markedly increases the probability works cited are intellectually distant. Other work found that scientists who briefly engage in conversation and are otherwise strangers become much more likely to collaborate and cite one another, but only if they are not from the same field (Lane et al., 2020). Finally, a recent survey of scientific authors about their citation behavior finds that cited researchers who reside within the same institution, but not the same department, are most likely to influence their research decisions (Duede et al., 2021). Universities and institutes bring intellectually distant researchers—insiders and outsiders—into close, physical proximity in a way that facilitates the very conversations that most profoundly catalyze future work.

Other quantitative work has focused attention on how science as a system searches for new ideas (Shi et al., 2015). Using research citations and compared content from prior work, a recent paper demonstrates that articles over the long 20th Century historically increased their intellectual reach by referencing prior work more distant to their own through the 1960s and stably thereafter (Tandon, Fortunato and Evans 2021). In this way, modern science may not be best portrayed as increasing in specialization (Jones, 2009; Price, 1965; Price and De Solla Price, 1963) and disunity (Dupré, 1993; Fodor, 1974; Stump and Galison, 1996), but as a reservoir of diversity. Increasingly scientists use this diversity as a source of foreign "patterns, theories, concepts and ideas from other fields to solve their own mysteries" (Tandon, Fortunato and Evans 2021).

Another recent paper reveals how across the life and physical sciences the degree to which a research article disrupts past research—the degree to which it violates our best predictions based on models built from prior research—is itself the best available predictor of outsized success in terms of awards and citations. These surprising successes systematically emerge across, rather than within researchers; most commonly when those in one field surprisingly publish problem-solving results to audiences in a distant other (Shi and Evans, 2019). Together these findings trace an undertheorized process through which punctuated scientific changes occur through collective abduction where parties to the conversation come from different fields with distinct intellectual, cultural and material resources.

**Sustainable Surprise**

From these observations it follows that a critical ingredient in disciplinary expansion is the existence of diverse, alternative theories, methods, logics, and empirical patterns. Recent analyses observe that interdisciplinary research tends to be underfunded relative to work within disciplines, largely because disciplinary insiders are those most likely called upon to peer review manuscripts, sit on funding panels (Bromham et al., 2016; Teplitskiy et al., 2018), and preside over award committees (Szell et al., 2018; Zivin et al., 2019). This stands in contrast to the observation that work at disciplinary intersections typically receives higher citations and is accorded outsized importance because it may surprise one or more of the communities whose expertise contributed to the work (Larivière et al., 2015; Larivière and Gingras, 2010; Leahey et al., 2017; Uzzi et al., 2013). This discrepancy, unearthed by science policy research, has led to an almost unending refrain extolling science policy makers to promote interdisciplinarity, despite the difficulty of implementation in the face of well-organized disciplines. But what if calls to interdisciplinarity succeeded?

A similar question was asked by structural sociologists who sought to understand whether the advantage of social agents who bridged structural holes in a social network could be sustained if everyone sought to do the same. The answer is no. The benefits of brokering the network would become evenly distributed (Burt and Hobart W Williams Professor of Sociology and Strategy Graduate School of Business Ronald S Burt, 2005: 230–233; Buskens and van de Rijt, 2008). This sounds like a socially positive outcome. The more who bridge social and cultural divides, the more ideas and cultural objects spread unrestrained throughout the network. On the surface, this might appear positive for science, where isolated disciplines lock up findings and theory in jargon that could benefit all if liberated. In the short term, this benefit might be realized by more interdisciplinary forays. But in the long term, ubiquitous conversation and networking between fields would consume the very difference between fields that is exploited fruitfully by interdisciplinarity. Recent work has explored the tightly coupled relationship between dense connections and contraction in the diversity of topics explored within 21st Century physics and science as a whole (Li et al., 2020). Dense, integrating collaboration between researchers and between university departments is systematically associated with a reduction in the space of topics physicists study. Complex, self-reinforcing causes underlie this association—more interdepartmental and interdisciplinary collaboration correlates conversations across science, while greater shared focus increases the ease and naturalness of collaborations.

The unlikely possibility of ubiquitous interdisciplinarity highlights an essential tension between the supply and demand for difference. If we want to shake loose the diversity of ideas for all of science we should unleash the potential for maximum abduction through interdisciplinarity, but at the cost of reducing longer-term differences between fields that serve as reservoirs for future abduction. To cultivate the capacity for science to generate sustainable surprises and change, paradoxically, we must protect the independence and separation of disciplines amidst calls for interdisciplinary convergence. Perhaps disciplines do not need more loyal patriots than they already raise, but despite explosive growth in the global human population, human languages go extinct at an accelerating pace with 50% on the cusp of being forgotten (Malone, 2005). In a shrinking world of digital communication and ubiquitous connection, here we fortify a new justification for disciplines by articulating the values of disciplinary difference and separation, which enable the sustainable bridging needed to generate abductive breakthroughs where the ideas and experiences from one field can resolve the anomalies of another. More disciplines with distinct, heterogeneous commitments protect the carrying capacity for sustainable diversity of ideas across science and scholarship.

How do fields of scientific ideas become and remain separate from one another? Some separations arise from forces beyond science. For instance, geo-political boundaries separate fields from one another. Because the vast majority of global research is supported by taxation within national innovation systems (Lundvall, 2010; Nelson and Rosenberg, 1993; Niosi et al., 1993), differences in national priorities maintain differences in science. Moreover, political demands to spread the investments in science broadly, even representatively across regions or universities, reduce the collapse of science into single, multi-disciplinary hubs in much of the world. Other separations emerge from the science itself. Scientific jargon is forged within fields of focused investigation to accelerate internal communication, but it limits how efficiently ideas can be transmitted across field boundaries (Vilhena et al., 2014). Finally, meaningful separation arises from the ontological and metaphysical characteristics of the natural world. Studying economies, the rise and fall of civilizations, tiger hunting patterns, or the works of Chaucer in terms of atoms and electrons would be "clearly insane" (Wallace, 2012).

These patterns reflect some of the same processes by which genetic populations speciate, and with many of the same consequences. If populations separate too far for too long then genetic communication (e.g., sex for animals) becomes impossible. But if they remain too close and in constant communication, interbreeding cannot confer hybrid vigor because minimal new genetic information is passed (Chen, 2010). Sustainable genetic diversity that supports evolution robust to environmental change can be bolstered not only by diversity of genes within a single genetic population. Multiple populations, each facing different environmental pressures and sufficiently separated, do not merely protect existing diversity, but evolve new traits that become relevant to one another with shifts in the environment.

Differences between disciplines make it difficult for the same individual to fully participate in certain combinations of scientific fields. Fields may promote logically inconsistent or incommensurable scientific processes or modes of explanation. Even though economics and sociology are close and often focus on the same phenomenon, such as income inequality and innovation, they often assume mutually exclusive models of human behavior, caricatured as *homo economicus* and *homo sociologicus (Hirsch et al., 1990)*, undersocialized and oversocialized humans whose choices are either driven by individual abstract utility or determined by cultures and context. These cartoons elude the realities of either nuanced population of perspectives; their orientations differ sufficiently that they come in theoretical conflict and require complex discursive reasoning to align (e.g., Granovetter, 1985). Similar conflicts arise between molecular biology and ecology, or particle and condensed matter physics (Li et al., 2020), which contrast a focus on the singular monads of existence—individuals, molecules, particles—with a focus on the complexity of interactions between them. Theoretical incommensurability can often lead to computational intractability. To mentally compute the implications of multiple, conflicting or even irrelevant perspectives simultaneously requires competing cognitive skills. Further, social and cultural barriers are often erected between neighboring fields in conflict, instigated to protect and insulate theoretical or methodological commitments and which drive toward their elaboration (see Kang and Evans, 2020 for the case of qualitative versus quantitative science studies).

For these incommensurabilities, it is much more likely for scientific contradiction and the possibility of abduction to occur across people across fields than within them. It also reinforces the importance of bringing back philosophy and the structure of scientific claims to the social studies of science as a locus of creative and conflictual social action. Bringing philosophy back in gets us closer to the constrained practice of making claims acceptable to scientists within a field, but as motivated by intuitions and patterns far beyond it.

Science policy and awards often reinforce the boundaries between fields as driven by insiders. This suggests another reason to bolster field boundaries—so that they can be productively, strategically and sustainably broken in future. This justification for science policy suggests how not trying to solve a problem with solutions from other fields paradoxically grows the infrastructure relevant to solving it. Sustaining reservoirs of unrelated scientific patterns allows them to draw upon each other at the limits of their reason.

**Scholarly Opportunity**

How then might we sustain or improve a scientific system that thrives on homogeneity but advances on heterogeneity? Our work lends support to the role of biography in intellectual history (Richards, 2017). While biography has typically supported the dubious idea of the lone genius, when done well, it documents the intimate interactions that lead to debates and findings that move fields. Consider the intellectual interaction between Hobbes and Boyle that carried with them centuries of commitments passed down from the Scholastics and Aristotle on one hand, and the new empiricism of experimental, natural philosophy on the other (Shapin and Schaffer, 2011). These are the interactions that typify social abductions and speak to the unique and enduring power of qualitative, historiographic analysis. They are suggestive of the importance of and a need for a genre of "conversation".

Digital traces of conversation can also be assembled with modern computation. A vast amount of evidence for how scientists and scholars think, hypothesize, approach and solve problems is latent in the enormous, now almost fully digitized, republic of letters. Each journal article, conference paper, book, pre-print or slide-deck contains part of the story of its own production, encoded in its roster of participating authors, discussion of prior art, methods, as well as citations to prior literature (Evans and Foster, 2011). Large-scale analysis of digitized scientific texts has already proven fertile for understanding research problem selection and novelty in science as well as for understanding how the institutions of science, current incentives, and publishing institutions influence the balance of disciplinary tradition characteristic of insiders, and novelty in research characteristic of surprising abductions (Fortunato et al., 2018). Analysis of millions of research publications across biomedicine has shown that high-risk innovation strategies are rare and reflect a growing focus on established knowledge and the recombination of insider concepts and entities (Rzhetsky et al., 2015; Shi et al., 2015). This lends empirical support to the dilemma outlined above. Breakthrough science is rare, in part, because insider science is both necessary for progress but simultaneously limits what can be proposed.

This suggests that our characterization of insiders accurately captures some of the limitations imposed by disciplinary *habitus* in the resolution of anomaly. The furious recombination of established knowledge, concepts, and entities within a field demonstrates a lack of novel concepts and entities to apply. This is fertile ground for anomaly. Yet, novelties from another field represent raw material for successful abductions that resolve anomalies. Already, we may be able to predict future abductions conceived as social syllogisms with a social calculus over these large scale data (Sourati and Evans, 2021). Moreover, we see an enormous, empirical opportunity to theorize how science could be done if outsiders with appropriate patterns of experience and expectations were effectively matched to insiders facing frustrating and stubborn anomalies. This also poses an empirical opportunity to verify and extend past conjectures concerning the nature of scientific transformation.

Following the idea of abduction as a social logic of inference, recent work argues empirically that the most remarkable successes in science materialize across unlikely combinations of scientists and scientific contexts rather than within homogenous contexts and between homogenous researchers (Shi et al., 2019; Shi and Evans, 2019). This work locates the emergence of the most surprising discoveries in science not within researchers holding diverse experiences, or even teams with diverse members, but in expeditions of researchers from outside a field, traveling into a field's problems and presenting their solutions in its conferences and journals.

This suggests how much more could be done if we zoom in to examine the logic of scientific claims, not merely juxtapositions of scientific concepts in detail. Claims making is a critical component of scientific action, and we can now begin to link it more directly with cultural patterns and milieu. New methods from machine learning have made it possible to turn unstructured historical text into structured text representations that index the cultural associations characteristic of particular times and places (Kozlowski et al., 2019). This could allow us to generalize the Foreman thesis (Foreman, 1971; Jammer and Merzbacher, 1967) into a large-scale empirical program. Foreman persuasively argued that the cultural rejection of determinism and materialism prevalent in the early 20th century Weimar Germany led to a conceptual rejection of these qualities characteristic of quantum physics being developed in that time and place. Blaming causal realism in politics by physicists who not only read, but also wrote about it, opened a space for the acceptance of theory that embraced the unpredictable behavior of quanta. More generally, a Foreman research program would identify which ambient cultural associations and models suggest particular scientific associations and influence particular scientific claims to seem more natural or unnatural—to feel more supported and obvious or anomalous and improbable by members of that culture.

Another empirical opportunity involves linkage of publication data to biographical and historical records, scientific lineages, and institutional records. This would enable science studies researchers to trace subjectivities along and across intellectual boundaries. It could allow the linking of these subjectivities with precise approaches to claims-making in order to identify how the landscape of claims and logic—identities, similarities and incommensurabilities—represent another critical landscape on which scientific action unfolds. It might further enable identification of specific scientific perspectives *relative to* one another in order to anticipate, isolate, and expose insider / outsider boundaries where abductions by social syllogism have proven and could prove most generative.

Finally, big data and machine learning models create the possibility to build tightly fitting models of big scholarly data, which enable the prediction of conventional and punctuated evolution in science. From a statistical perspective, these models are often overfit and may not generalize, but if they are constructed over the entire population of relevant scientific claims and claimants, they can trace expectations by scientists, which are themselves "overfit" to their personal histories and experiences. With tight, data-driven models of situated scientific expectations, we can begin to model scientific surprise in a way that resonates with Pierce's own semiotic characterization of the relationship between signs, objects, and the bridges or interpretants that link the two in human meaning-making (Danesi, 2004; Muirhead, 1932). This third aspect traces the process by which human minds semiotically mould around objects not only through experience, but habituated socialization.[12] This could enable us to directly model collective Piercian interpretants, allowing for large-scale analysis and simulation of many forms of phenomenological surprise and abduction.

# Discussion

While the role of abductive inference in science has not garnered as much attention as inductive and deductive processes, abduction has, nevertheless, been of interest to philosophers of science for quite some time (Peirce, 1957) and has more recently been explored by qualitative sociologists (Abbott, 2004; Stefan Timmermans and Iddo Tavory, 2012) as a means of accelerating discovery. However, virtually all of the literature on abduction has focused on the process of inference taking place in the head of a singular reasoning subject. Indeed, when considered as a singular process of "inspiration", abduction can easily be characterized as a process that takes place in a black box, a mind driving question discovery as a function of individual experts. As a result, abduction in the philosophy of science has cleaved to epistemological questions concerning the justificatory credentials of abductive inference, leaving its role in discovery to sociologists or psychologists. On the other hand, the criticality of social interaction in science is constitutionally obvious to social scientists and science studies scholars, but with philosophers and logic in science scant at the science studies buffet, there has been little appetite for articulating the social role that multiple minds play in supplying the necessary premises for successful abductive inferences. As a result, abduction has remained a scholarly heuristic that periodically rises to prominence in social science as a guide to thinking better (Glaser and Strauss, 1967; Stefan Timmermans and Iddo Tavory, 2012; Tavory and Timmermans, 2014). By bringing the concerns of philosophy back to the science studies table and recognizing the scientific arguments over which philosophers obsess as critical modes of scientific practice, we demonstrate the contribution that social interaction across epistemic cultures can make to the plausibility of abduction 'in the wild' of science (Hutchinson, 1995) and its relevance for the punctuated evolution of scientific ideas.

---

[12] Pierce described thirdness as "synthetic consciousness driven by the sense of learning, thought, memory and habit" (Muirhead, 1932).

In this paper, we have argued for two points concerning the nature of scientific expertise which explain the transformative role that abduction plays in science when conceived not as taking place in the mind of an individual, but as a social process taking place across individuals. The first point is that, for a given scientific actor, the intellectual resources that make her most sensitive to or surprised by an anomaly are her exposures and commitments to the theoretical principles that govern her discipline. What constitutes her expertise is precisely this sensitivity to what should and should not happen under certain conditions. Our second point is that expertise may also make her significantly less aware of the foreign patterns of thought, systems of claims, or empirical observations relevant to the production of theoretically innovative solutions to empirical events that strike her as anomalous.

Unlike ampliative inferences such as induction from which truly novel, generalized claims can be posited directly from the experience of a single observer, abductions cannot generate new propositions. Rather, an abductive inference requires a search through the background knowledge in the abducting agent for a proposition (the explanation) that best satisfies what context demands. An anomaly is anomalous precisely because it is inconsistent with the propositions an agent holds, but when cast as a social inference, the "dilemma" is resolved. Advancing the claim that abduction in science constitutes a social process describes not only a relationship between individual scientists, but also brokers a renewed one between the social studies and philosophy of science.

Advancing the idea that abduction is carried through conversations enables the system of science to abduct novel discoveries even and especially when individual scientific agents cannot. This underscores the importance of continued disciplinary work for successful interdisciplinary outcomes, which rely on disciplinary certainties from which they can depart. It is precisely the parochial, separatist and even elitist impulse in scientific fields that make them relevant to other disciplines over the long term, suggesting the importance for science studies to document, and science policy to protect disciplinary diversity for sustainable innovation.